\documentclass[final]{aipproc}
\usepackage{amsmath,amssymb}
\layoutstyle{6x9}

\begin{document}

\title{MFV and the SUSY CP Problem}

\classification{
11.30.Er, % C, P, T and all that
11.30.Hv, % flavour symmetries
12.60.Jv, % SUSY
31.30.Jp  % electron EDM
}
\keywords{Supersymmetry, Minimal Flavour Violation, CP violation, electric dipole moments}

% \author{Paride Paradisi}{
%   address={Physik-Department, Technische Universit\"at M\"unchen, 85748 Garching, Germany}
% }

\author{David M. Straub}{
  address={Physik-Department, Technische Universit\"at M\"unchen, 85748 Garching, Germany}
}

\begin{abstract}
The principle of Minimal Flavour Violation (MFV) provides a natural solution to the SUSY flavour problem, but does not solve the SUSY CP problem as it allows for the presence of new CP-violating phases. If the MFV principle is generalized by the assumption of CP conservation in the limit of flavour blindness, the SUSY CP problem can be solved, predicting at the same time interesting CP violating phenomenology.
% An important issue is the RG evolution of phases if this assumption is imposed at high energies.
\end{abstract}

\maketitle

%%%%%%%%%%%%%%%%%%%%%%%%%%%%%%%%%%%%%%%%%%%%
%% MAINMATTER
%%%%%%%%%%%%%%%%%%%%%%%%%%%%%%%%%%%%%%%%%%%%

\section{Introduction}

Generic SUSY theories introduce new sources of flavour and CP violation, which are strongly constrained by the excellent agreement of flavour physics data with the SM expectations. An elegant way to solve this SUSY flavour problem is represented by the principle of Minimal Flavour Violation (MFV) \cite{Buras:2000dm,D'Ambrosio:2002ex}. The MFV ansatz makes use of the global flavour symmetry $G_\text{f} \sim SU(3)^5$ present in the SM in the absence of Yukawa couplings. By formally promoting the Yukawas to spurions of the broken $G_\text{f}$ and demanding that all couplings are invariant under $G_\text{f}$, flavour violation arises in MFV theories only through the Yukawa couplings, and is thus essentially described by the CKM matrix. However, the MFV symmetry principle does not by itself provide a solution to the SUSY CP problem \cite{Colangelo:2008qp,Mercolli:2009ns}. In particular, flavour blind phases, such as the phase of the $\mu$ term or of the diagonal trilinear parameters $A_I$ in the MSSM, are not forbidden. Thus, an extra assumption or an extra mechanism is required to suppress these phases, which are strongly constrained e.g. by the non-observation of the electric dipole moments (EDMs) of the electron or neutron.

Instead of making the extreme assumption that the CKM phase is the only source of CPV in the MFV MSSM \cite{D'Ambrosio:2002ex}, an alternative approach was suggested in \cite{Paradisi:2009ey}: the soft SUSY breaking sector is assumed to be CP conserving only in the limit of flavour blindness (i.e. in the limit where the soft masses are universal and the trilinear couplings are proportional to the respective Yukawa couplings), but CP is assumed to be violated by the MFV-compatible terms breaking the flavour blindness. This is reminiscent of the approach frequently adhered to in studies of SUSY flavour models, where the theory is assumed to be CP conserving in the flavour symmetric limit, with CP violation only arising from terms breaking the flavour symmetry.

This approach can be illustrated by the squark mass matrices and trilinear couplings, which can be written in the case of MFV as an expansion in powers of Yukawa couplings,
\begin{equation}
\mathbf{m}_{Q}^{2} = m_{Q}^{2}
\left[
\mathbf{1}+r_{1}\mathbf{Y}_{u}^{\dagger}\mathbf{Y}_{u}+r_{2}\mathbf{Y}_{d}^{\dagger}\mathbf{Y}_{d}+
(c_{1} \mathbf{Y}_{d}^{\dagger}\mathbf{Y}_{d}\mathbf{Y}_{u}^{\dagger}\mathbf{Y}_{u}+\mathrm{h.c.})
\right]\;,
\label{MFV:mQ}
\end{equation}
\begin{equation}
\mathbf{A}^{U} =
A_{U}\mathbf{Y}_{u}\left( \mathbf{1} + c_{3}\mathbf{Y}_{d}^{\dagger}\mathbf{Y}_{d}+
c_{4}\mathbf{Y}_{u}^{\dagger}\mathbf{Y}_{u} +
c_{5}\mathbf{Y}_{d}^{\dagger}\mathbf{Y}_{d}\mathbf{Y}_{u}^{\dagger}\mathbf{Y}_{u}+
c_{6}\mathbf{Y}_{u}^{\dagger}\mathbf{Y}_{u}\mathbf{Y}_{d}^{\dagger}\mathbf{Y}_{d}
\right)\;,
\label{MFV:AU}
\end{equation}
and similarly for the remaining couplings in the sfermion sector. The parameters $r_i$ have to be real due to the hermiticity of the $\mathbf m_i^2$, while the $c_i$ parameters can be complex. The assumption of CP conservation in the limit of flavour blindness now amounts to assuming $A_U$ etc. to be real, as well as the $\mu$ parameter and the gaugino masses, while all the $c_i$ parameters are allowed to be complex.
The question to address now is whether, under these assumptions, $O(1)$ phases in the $c_i$ are compatible with the data on CPV observables. Crucially, however, this question depends on the scale at which these assumptions are imposed. Therefore, let us answer it in turn for a scenario defined at low energies, and for a scenario defined at the GUT scale.

\section{Low scale MFV scenario}

If the assumption that CPV in the MFV MSSM arises only through terms breaking the flavour blindness is imposed at the EW scale, then the $\mu$ term and the gaugino masses are real at low energies and CPV beyond the CKM phase arises only from the $c_i$ coefficients in the $A$ terms (neglecting the highly suppressed $c_i$ in the squark mass matrices).

In particular, looking at eq. (\ref{MFV:AU}), it is clear that the imaginary parts of the $A$ terms, which arise through the $c_i$, scale with the cube of the fermion masses and are thus strongly hierarchical: $\text{Im}\, A_t\gg \text{Im}\, A_{c,u}$, and accordingly $\text{Im}\, A_b\gg\text{Im}\,
A_{s,d}$ and $\text{Im}\, A_{\tau}\gg \text{Im}\, A_{\mu,e}$. Since the most stringent constraint on this scenario stems from the upper bounds on EDMs related to first generation fermions and since the one-loop contributions to these EDMs are proportional to the imaginary parts of the first generation $A$ terms, this hierarchy strongly suppresses the dangerous one-loop contributions to EDMs.

Then, the dominant EDM contribution arises through a two-loop Barr-Zee type diagram proportional to the stop trilinear coupling. While these diagrams contribute both to lepton EDMs and to quark (C)EDMs, the most constraining observable turns out to be the electron EDM (measured through the Thallium EDM).

Numerically, the phenomenological viability of the low-scale scenario can thus be assessed by concentrating on the predictions for the electron EDM, and on phases in the two dominant terms in the MFV expansion of $\mathbf A_U$. The left panel of fig. \ref{edm_mfv} shows the results of a scan assuming a common SUSY mass and complex $A$ terms, arising from the terms proportional to $c_3$ and $c_4$ in eq. (\ref{MFV:AU}), assuming purely imaginary $c_{3,4}$ to be fair. In both cases, the EDM is well under control, but can reach experimentally visible levels.

\section{GUT scale MFV scenario}

If the MFV expansion holds at the GUT scale, it remains valid down to low energies \cite{Paradisi:2008qh,Colangelo:2008qp}; however, this is not necessarily true for the ansatz that CPV arises only from terms breaking the flavour blindness, since phases in the complex parameters in eq. (\ref{MFV:AU}) etc. can be generated through the RG evolution from the GUT scale to the EW scale.

In particular, the imaginary parts of the first generation $A$ terms, which are highly suppressed by their cubic dependence on the fermion masses at the input scale (as discussed in the previous section), can become sizable. This can be illustrated by an approximate solution to the RG equation of the first-generation up-squark trilinear $A_u$,
\begin{equation}
A_u(m_Z) \approx
A_u(m_G)- 0.41 y_t^2 A_U + 0.03 y_b^2 A_D
-\left( 0.05 y_t^2 y_b^2 c_{3} + 0.11 y_t^4 c_{4}\right) A_U - 2.8m_{1/2}.
\label{Au_fit}
\end{equation}
The ansatz described in the introduction amounts to assuming $m_{1/2}$ and $A_{U,D}$ real and $c_{3,4}$ complex, while the GUT scale value $A_u(m_G)$ is nearly real due to the cubic suppression discussed above. As can be seen from (\ref{Au_fit}), a sizable imaginary part can be induced in $A_u(m_Z)$ through the terms proportional to $c_{3,4}$, with potentially dangerous effects for EDMs. However, this effect competes with a large real contribution to $A_u$ induced by $SU(3)$ interactions and proportional to $m_{1/2}$.

Similarly, the stop trilinear coupling, which drives the dominant two-loop contributions to the EDMs, can be expressed at low energies as
\begin{equation}
\begin{split}
A_t(m_Z) \approx
A_t(m_G)- 0.81 y_t^2 A_U
- 0.09 y_b^2 A_D 
+ \left( 0.04 y_t^2 y_b^2 c_{3} + 0.10 y_t^4 c_{4}\right) A_U \\
- \left( 0.03 y_t^2 y_b^2 c_{7} + 0.01 y_b^4 c_{8}\right) A_D - 2.2 m_{1/2}.
\label{At_fit}
\end{split}
\end{equation}
In this case, already the value at the GUT scale $A_t(m_G)$ can have a sizeable phase, since it is not suppressed by small Yukawa couplings as can be seen from eq. (\ref{MFV:AU}). However, this phase is strongly reduced by the last term of (\ref{At_fit}), since the assumption of CP conservation in the limit of flavour blindness dictates $m_{1/2}$ to be real.

A more subtle, but nevertheless crucial, effect regards the phase of the $\mu$ term. Since the basic ansatz requires $\mu$ to be real at the GUT scale, the well-known RG invariance of the phase of $\mu$ implies that $\mu$ remains real at low energies. However, one now has to worry about an assumption frequently made about a related parameter: the SUSY breaking $B\mu$ parameter. From a phenomenological perspective, the value of $B\mu$ at the EW scale is fixed by the EWSB conditions, and it is required to be real to obtain real Higgs VEVs. However, the phase of $B\mu$ is not RG invariant; on the contrary, it can pick up a large phase in the presence of complex $A$ terms. Thus, the phenomenologically necessary (and often tacit) assumption that $B\mu$ is real at the EW scale requires, in the presence of complex $A$ terms, a complex $B\mu$ term at the GUT scale. While this might be acceptable in phenomenologically motivated scenarios like the CMSSM, it clashes with our assumption of a CP conserving theory in the absence of flavour breaking.

The solution to this conundrum can be obtained by a change of basis. Assuming $\mu$ and $B\mu$ to start out real at the GUT scale, $\mu$ will remain real and $B\mu$ become complex at the EW scale. By exploiting the $U(1)_\text{PQ}$ phase transformation on the Higgs superfields, which amounts to an opposite phase shift in $\mu$ and $B\mu$, $B\mu$ can be made real, as required by EWSB, but then $\mu$ will be complex!

As a consequence of this mechanism, the $\mu$ term acquires an effective CPV phase through RG effects in our scenario, giving rise to additional one-loop contributions to EDMs. It should be stressed that this mechanism is not restricted to MFV models, but is at work in all models predicting complex $A$ terms, but real $\mu$ and $B\mu$ terms at the GUT scale.

For the numerical assessment of its phenomenological viability, also in the GUT scale scenario it is most instructive to restrict oneself to phases only in $\mathbf A_U$ at the input scale, since the RG effects are driven most strongly by the stop trilinear.

\begin{figure}[tb]
\includegraphics[width=0.45\textwidth]{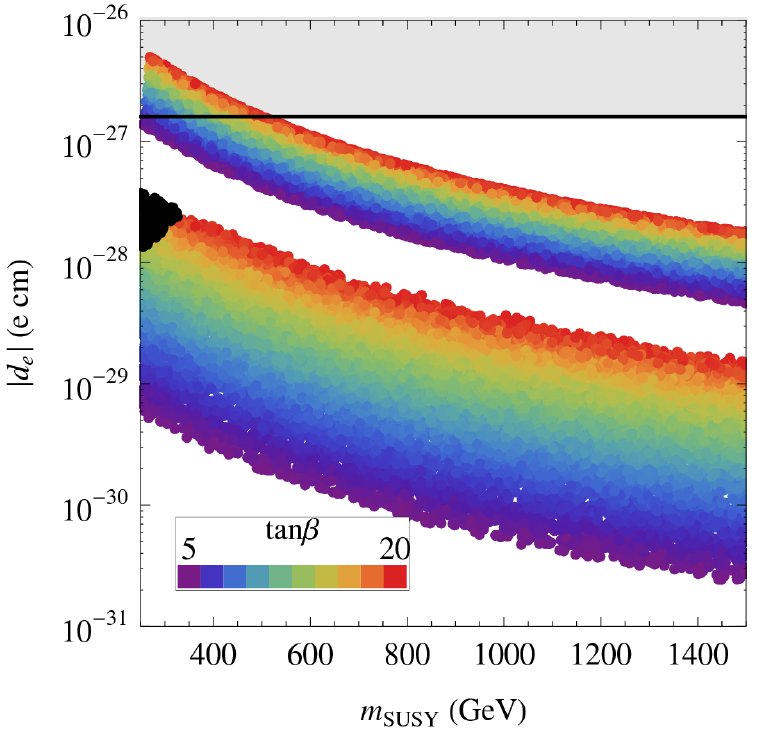}
\qquad
\includegraphics[width=0.45\textwidth]{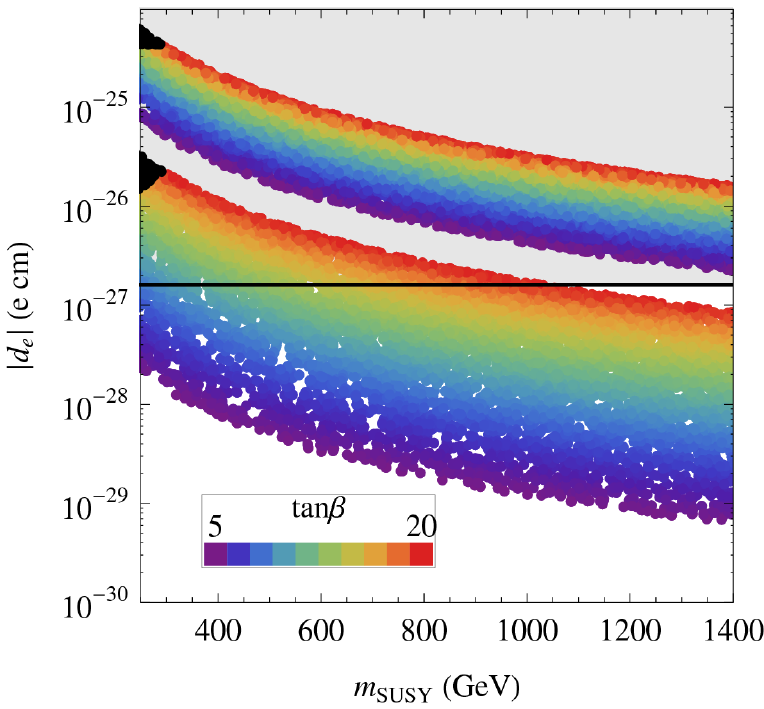}
\caption{Predictions for the electron EDM in MFV frameworks defined at the EW scale (left) and at the GUT scale (right) assuming the boundary conditions
$\mathbf{A}^{U}=A_{U}\mathbf{Y}_{u}(\mathbf{1}+c_4\mathbf{Y}_{u}^{\dagger}\mathbf{Y}_{u})$
with $c_4=i$ (upper bands) and
$\mathbf{A}^{U}=A_{U}\mathbf{Y}_{u}(\mathbf{1}+c_3\mathbf{Y}_{d}^{\dagger}\mathbf{Y}_{d})$ 
setting $c_3=i$ (lower bands).
In the EW-scale scenario, all dimensionful SUSY breaking parameters where assumed to be degenerate at $m_\text{SUSY}$. In the GUT-scale scenario, $A_U=A_0=m_0=m_{1/2}\equiv m_\text{SUSY}$ was assumed.}
\label{edm_mfv}
\end{figure}

The results of a scan where the same MFV terms in $\mathbf A_U$ as in the previous section (this time at the GUT scale) have been switched on and assumed to be purely imaginary are shown in the right panel of fig. \ref{edm_mfv}. The scenario with $c_4=i$ is ruled out for all values of $\tan\beta$ even for a sizable SUSY scale. The scenario with $c_3=i$, however, is allowed at small to intermediate $\tan\beta$ even for a quite low SUSY scale.

As a final comment, while this study was limited to the EDMs, being the most constraining observables, these scenarios also have a rich and interesting $B$ physics phenomenology, as recently studied in \cite{Altmannshofer:2009ne}.

\begin{theacknowledgments}
I would like to thank Paride Paradisi for the pleasant and fruitful collaboration. This work has been supported in part by the DFG cluster of excellence ``Origin and Structure of the Universe''.
\end{theacknowledgments}

\bibliographystyle{aipproc}
\bibliography{mfvcp}

\end{document}